\documentstyle[12pt,aasms4]{article}

\received{1998 November 25}
\accepted{1999 February 3}
\slugcomment{to appear in {\it The Astrophysical Journal}}

\lefthead{Kraemer, Ho, Crenshaw, Shields, \& Filippenko}
\righthead{The Emission-line Gas of NGC 4395}

\begin{document}

\title{Physical Conditions in the Emission-Line Gas \\
in the Extremely Low-Luminosity Seyfert Nucleus of NGC 4395
\altaffilmark{1,2}}

\author{Steven B. Kraemer\altaffilmark{3},
Luis C. Ho\altaffilmark{4},
D. Michael Crenshaw\altaffilmark{3},
Joseph C. Shields\altaffilmark{5},
\& Alexei V. Filippenko\altaffilmark{6},
}

\altaffiltext{1}{Based in part on observations made with the NASA/ESA {\it Hubble Space 
Telescope}. STScI is operated by the Association of Universities for Research in 
Astronomy, Inc. under the NASA contract NAS5-26555. }

\altaffiltext{2}{Based in part on observations made with the 
Infrared Space Observatory, an ESA project with instruments funded by ESA
Member States with the participation of ISAS and NASA.} 

\altaffiltext{3}{Catholic University of America,
NASA/Goddard Space Flight Center, Code 681,
Greenbelt, MD  20771; stiskraemer@yancey.gsfc.nasa.gov, 
crenshaw@buckeye.gsfc.nasa.gov.}

\altaffiltext{4}{Carnegie Observatories, 813 Santa Barbara Street,
Pasedena, CA 91101-1292; lho@ociw.edu.}

\altaffiltext{5}{Department of Physics \& Astronomy, Ohio University,
Athens, OH 45701-2979;
shields@helios.phy.ohiou.edu.}

\altaffiltext{6}{Department of Astronomy, University of California, Berkeley,
CA 94720-3411; alex@astro.berkeley.edu.}

\begin{abstract}

  We have combined {\it Hubble Space Telescope}/Faint Object Spectrograph,
  ground-based, and {\it Infrared Space Observatory} spectra
  of the nucleus of NGC 4395, the least luminous and nearest known type 1 
  Seyfert galaxy. The spectra show emission lines from a wide 
  range of ionization states and critical densities.
  We have generated multicomponent photoionization models
  of both the broad and narrow emission-line regions (BLR and NLR) to
  investigate the physical conditions in the emission-line gas and
  test the proposition that the source of ionization is the non-stellar
  continuum radiation emitted by the central source. We show that, with
  a minimum of free parameters, the model predictions match the observed 
  emission-line intensity ratios quite well. The elemental abundances appear to be 
  subsolar, with even greater underabundance of nitrogen.
  From the size of the BLR
  predicted by the models, we estimate a central mass of a few x 
  10$^{5}$ M$_\odot$, in reasonable agreement with estimates from the
  stellar kinematics. Finally, our results suggest that the covering factor of 
  the emission-line gas is close to unity, and that the observed UV to X-ray
  continuum is absorbed by intervening NLR gas. We argue that a
  high covering factor is responsible for the apparent flattening of the Baldwin relation
  in low-luminosity active galactic nuclei. 

\end{abstract}


\section{Introduction}

 NGC 4395 is a nearby (d $\approx$ 2.6 Mpc; Rowan-Robinson 1985), late-type 
 (Sd IV) dwarf galaxy which, in general characteristics, bears a 
 resemblance to the Large Magellanic Cloud (LMC). 
 Both show a weak bar-like structure, bright H~II regions, and possess 
 subsolar abundances of heavy elements (for LMC, see Russell \& Dopita 1992; 
 for NGC 4395, see Roy et al. 
 1996). However, there is a bright point source very near the center of 
 symmetry of NGC 4395. Optical and UV spectra of this source reveal
 a featureless continuum and strong emission lines from a wide range
 of ionization states and critical densities. 
   
 Although the bolometric luminosity of the central knot ($\sim$ 1.5 x 10$^{40}$
 ergs s$^{-1}$) could be produced by massive stars, there is ample evidence that
 NGC 4395 harbors a dwarf Seyfert 1 nucleus. This is discussed in
 detail in two previous papers (Filippenko \& Sargent 1989, hereafter FS89; 
 Filippenko, Ho, \& Sargent 1993, hereafter FHS93), and here we will only
 summarize the evidence. First, the optical spectrum shows coronal lines, such as 
 [Ne~V] $\lambda\lambda$3346,~3426
 and [Fe~VII] $\lambda$6087, which indicate the presence of a significant
 flux of photons with energies above 97 eV. Although it is possible that
 such energetic radiation can be produced by stars (Terlevich \& Melnick 
 1985), it is a straightforward consequence of the power-law continua
 associated with the active galactic nucleus (AGN) phenomenon. In fact, the optical to UV 
 continuum radiation emitted by the central source is best characterized 
 as a featureless power law and the optical continuum flux has been observed
 to vary by as much as 20\% on timescales $\sim$ 1 day (Lira, Lawrence, \& 
 Johnson 1999; but see Shields \& Filippenko 1992). The active nucleus appears as an 
 unresolved point source in the radio with a nonthermal spectrum 
 (Sramek 1992), and has been detected in the soft X-ray band (0.1 - 2.4 keV)
 with the {\it ROSAT}/PSPC (Moran et al. 1999). All this suggests
 that there is a compact energy source, analogous to Seyfert1 nuclei.
 Finally, the permitted lines have broad wings (FWHM $>$ 1000 km s$^{-1}$),
 which are absent from any of the forbidden lines,
 indicating dense (n$_{H}$ $>$ 10$^{9}$ cm$^{-3}$), fast moving
 gas, which is best explained by material close to the central
 mass/ionization source, and is one of the defining characteristics of
 a Seyfert 1 nucleus.

 It is fascinating that the nucleus of NGC 4395 appears to possess all of the 
 basic properties of type 1 Seyfert galaxies and QSOs; 
 the fundamental physics of the AGN phenomenon thus applies over at least
 eight orders of magnitude in luminosity! This object permits us to extend
 studies of luminosity dependent effects (e.g., the Baldwin effect) to the 
 lowest extreme. Furthermore,
 NGC 4395 is a very late-type galaxy, lacking a central bulge, unlike
 virtually every other known Seyfert galaxy. Finally, NGC 4395
 is close to us, which allows us to examine the host galaxy 
 with hitherto unprecedented spatial resolution.

 In this paper we examine in detail the physical conditions of the Seyfert nucleus
 of NGC 4395. In addition to the {\it Hubble Space Telescope}/Faint Object
 Spectrograph ({\it HST/FOS}) spectra presented in FHS93, 
 we include new ground-based optical spectra and IR emission-line spectra taken 
 with the {\it Infrared Space Observatory} ({\it ISO}). The combination of UV,
 optical, and IR data provides the widest range of emission-line diagnostics,
 as well as an opportunity to deredden the lines based on the He~II 
 recombination ratios.
 We will use multi-component photoionization models to match the
 dereddened emission-line ratios and probe the physical conditions
 in both the broad-line region (BLR) and narrow-line region (NLR) in this AGN.

\section{Observations and Data Analysis}

\subsection{UV and Optical Spectra}

We obtained UV spectra of NGC 4395 with {\it HST}/FOS using the G130H, G190H, and G270H gratings. The 
UV spectra cover the wavelength range 1150 -- 3300 \AA\ at a spectral 
resolution 
of $\lambda$/$\Delta\lambda$ $\approx$ 1000 (FWHM $=$ 1.2 -- 3.3 \AA~, over the 
FOS waveband), and are described in more detail 
in FHS93.
We obtained high-dispersion optical spectra during the course of an extensive 
spectroscopic survey of nearby galaxies conducted with the CCD Double 
Spectrograph (Oke \& Gunn 1982) on the Hale 5~m telescope at Palomar 
Observatory (see Filippenko \& Sargent 1985, and Ho, Filippenko, 
\& Sargent 1995, for details of the survey).  
The 1$''$ wide slit yielded spectral resolutions of FWHM $\approx$ 2.3 \AA\ 
and 1.6 \AA\ in the blue (4200--5100 \AA) and red (6200--6800 \AA) spectra,
respectively.
We also obtained spectra of moderate dispersion on several occasions at Lick 
Observatory using the Kast spectrograph (Miller \& Stone 1993) mounted on the 
Shane 3~m reflector.  These observations used a 2$''$ slit, and the 
wavelength coverage extended from near the atmospheric cutoff (3200 \AA) to 
almost 1 $\mu$m with a typical FWHM resolution of 5--8 \AA.  In all cases, we 
oriented the slit along the parallactic angle to minimize light losses due to 
atmospheric dispersion, and we observed featureless spectrophotometric standard 
stars to calibrate the relative fluxes of the spectra.  One-dimensional 
spectra were extracted using a 4$''$ window centered on the nucleus, with 
the sky background determined from regions along the slit far from the 
nucleus.  The reduction and calibration of the data followed conventional 
procedures for long-slit spectroscopy, such as those described in Ho et al. 
(1995).

The absolute fluxes of the UV spectra are uncertain, as evidenced by $\sim$30\%
discrepancies in the continuum fluxes of individual FOS spectra 
in the regions of overlap (possibly due to aperture
miscentering). We scaled the UV spectra to match each other and to 
match the optical spectra obtained on spectrophotometric nights. Based on a 
comparison of these optical spectra, we expect that the uncertainty in the 
absolute flux calibration is $\sim$20\%, in addition to the measurement errors 
that we quote later.

Figure 1 shows the UV spectrum and a representative optical spectrum, and 
demonstrates the large number of emission lines available for photoionization 
modeling.
The broad and narrow components of the permitted 
emission lines can most easily be seen in the C~IV $\lambda$1550 line,
as shown in Figure 2. The width of the broad C~IV $\lambda$1550 is $\sim$ 
5000 km s$^{-1}$, while broad H$\beta$ has a FWHM $\sim$ 1500 km s$^{-1}$
(also see Figure 2), which is evidence for stratification in the BLR. However,
as we will show in Section 3, the BLR gas is well represented by a single
set of physical parameters. Although the forbidden lines are
quite narrow ([O~III] $\lambda$5007 has a FWHM $\leq$ 
50 km s$^{-1}$, FS89), high-resolution (FWHM $=$ 8 km s$^{-1}$) optical spectra 
obtained with the Keck 10-m telescope (Filippenko \& Ho 1999) show
differences in the profiles of the low-ionization narrow lines,
such as [O~I] $\lambda$6300 and [S~II] $\lambda$6731, compared to 
higher ionization lines, such as [O~III] $\lambda$5007 and 
[S~III] $\lambda$6312. This is evidence that the NLR of NGC 4395 is 
stratified as well. Unlike the BLR, the NLR cannot be represented by a single
set of physical parameters, as we will discuss in Section 3.

\subsection{Mid-Infrared Spectra from {\it ISO}}

We acquired the IR spectra with the Short Wavelength Spectrometer (SWS)
using the AOT 2 mode to cover selected wavelength intervals for
measurement of specific lines, at an average spectral resolution 
of $\lambda$/$\Delta\lambda$ $\approx$ 1400. Spectral regions centered at the
redshifted ($z=$ 0.00101, Haynes et al. 1998) wavelengths of 
[S~IV] 10.51~$\mu$m and [S~III] 33.48~$\mu$m were observed
simultaneously with a total integration time of 10,400 s, while regions
appropriate for [Si~IX] 3.94~$\mu$m and [O~IV] 25.89~$\mu$m were observed simultaneously
with a total integration time of 18,600 s.  While observing the source, the
grating was stepped between readouts to produce a scan across the detectors
of 200 s duration, followed by measurement (with multiple readouts) of the 
dark current.

We reduced the spectra using the Interactive Analysis and {\it ISO}
Spectral Analysis version 1.3a software packages.  Standard
Processed Data generated by the calibration pipeline (version
5.2) were employed.  Dark current measurements were interactively
trimmed of large noise excursions using a 3$\sigma$ clipping routine,
and dark current estimates for each scan were generated from a mean of
dark measurements immediately bracketing the scan in time.  Individual
scans with discontinuous jumps in dark current levels were
interactively removed on a detector-by-detector basis, and the
measured dark current levels were subtracted from the remaining data.
Standard corrections were applied for detector response and flux
calibration, as well as wavelength correction to the heliocentric
frame.  For each detector, fluxes were averaged across the entire set
of observations for a given line; individual scans were then shifted
by a constant additive offset to bring their flux level into agreement
with this value, as a correction for residual errors in the dark
current removal.  An overall spectrum was then generated for each
detector by taking a median of individual scan spectra with 3$\sigma$
clipping.  After shifting again to a common mean flux level, the
spectra for individual detectors were then used to construct a
weighted average representing the final spectrum.  This process was
repeated with the data subdivided according to scan direction (up or
down) as a consistency check on the results.

The {\it ISO} spectra are shown in Figure 3. The plots display the full 
wavelength coverage of the data, corresponding to the wavelength coverage
per scan for each of the four lines. We have strong detections of the [S~IV] 
and [O~IV] lines, whereas the [Si~IX] and [S~III] lines are not visible.

\subsection{Measurements}

We measured the fluxes of the narrow lines that lack broad counterparts 
directly, whereas for severely blended lines like 
H$\alpha$ and [N~II] $\lambda\lambda$6548,~6583, we used the [O~III] 
$\lambda$5007 profile as a template to deblend the lines (see Crenshaw \& 
Peterson 1986). To measure the broad component of the permitted lines, we 
deblended and removed the narrow components, and determined the remaining flux 
above a local continuum. For some lines, the broad component was expected to be 
present but was too weak for a reasonable flux measurement.

We determined the reddening of the narrow emission lines from the He~II 
$\lambda$1640/$\lambda$4686 ratio and the Galactic reddening curve of Savage \& 
Mathis (1979). The He~II lines are due to recombination and are therefore 
essentially insensitive to temperature and density effects;
we adopt an intrinsic value of 7.2 for this ratio, which is consistent with our 
model values (Section 4). The observed He~II ratio 
is 6.0 $\pm$ 0.8, which yields a reddening of $E_{B-V}$ $=$ 0.05 $\pm$0.03 mag.
The reddening from our own Galaxy is $\leq$ 0.03 mag (Murphy et al. 1996).

Table 1 gives the observed and dereddened narrow-line ratios relative to 
narrow H$\beta$, and errors in the dereddened ratios. Table 2 gives the observed 
broad-line ratios, relative to broad H$\beta$, and dereddened broad-line ratios 
obtained by assuming the broad lines experience the same amount of reddening as 
the narrow lines.
We determined errors in the dereddened ratios from the sum in quadrature of the 
errors from three sources: photon noise, different reasonable continuum 
placements, and reddening.

\section{Photoionization Models}

As in our previous studies (e.g., Kraemer et al. 1998a), we have attempted
to keep the number of free parameters to a minimum by using the best 
available observational constraints and simplest of assumptions. For
example, the size of the emission-line region predicted by the
models could not exceed the upper limits from ground-based measurements
($\sim$ 13 pc, FHS93). The spectral energy distribution (SED) of the ionizing
continuum radiation was determined by simple fits to the observed fluxes in the UV and X-ray.
Since it is apparent that there are distinct broad
and narrow emission-line regions in the nucleus of NGC 4395, which cannot
be modeled assuming a single density, we initially assumed a single component each for 
the BLR and NLR gas. Additional components were added to improve the 
fit to the observed emission-line ratios, but only if they met the observational
constraints. 

  The details of the photoionization code are given in Kraemer (1985) and 
other papers (cf. Kraemer et al. 1994). Following convention, our
photoionization models are parameterized in terms of the
number of ionizing photons per hydrogen atom at the illuminated face of the
cloud, referred to as the ionization parameter:
 
\begin{equation}
U = {1\over{4\pi~D^2~n_H~c}}~ \int^{\infty}_{\nu_0} ~\frac{L_\nu}{h\nu}~d\nu,
\end{equation}
   where $L_{\nu}$ is the frequency dependent luminosity of the ionizing 
   continuum, $n_{H}$ is the number density of atomic hydrogen,
   $D$ is the distance between the cloud and the ionizing source, and 
   $h\nu_{0}$ =13.6 eV. 

   In the following subsections, we will discuss how values were assigned to
   the various input parameters.

\subsection{Elemental Abundances}

   Several studies have addressed the elemental abundances
   in the H~II regions within NGC 4395, and there is strong evidence
   that the O/H ratio is significantly subsolar (Vila-Costas \&
   Edmunds 1993; Roy et al. 1996; van Zee, Salzer, \& Haynes 1998). 
   Roy et al. (1996) derived a value of log(O/H) $=$ 
   $-$3.7 for an H~II region within $\sim$ 0.1 kpc of the nucleus,
   based on the [N~II]/[O~III] ratio and the semi-empirical
   calibration of Edmunds \& Pagel (1984).
   The accuracy of this estimate and its appropriateness for the nucleus
   are uncertain, however, since results reported by Roy et al. for other
   regions in this galaxy show a scatter spanning log(O/H) = $-$4.2 to
   $-$3.4, with no clear radial trend (see also Vila-Costas \& Edmunds 1993).
   Our test calculations suggest that the 
   [O~I] $\lambda$6300/H$\beta$ ratio is best matched with an abundance of
   at least log(O/H) = $-$3.5 ($\sim$ 1/2 solar; Grevesse \& Anders 1989), 
   in plausible agreement with the extranuclear results, and therefore
   we adopt this value for our analysis.
   
   The evidence is strong that the N/O abundance ratio in NGC 4395 is also 
   subsolar. Estimates based on optical [N~II]/[O~II] line ratios for
   H~II regions in this galaxy generally fall in the range of log(N/O)
   $\approx$ $-$1.5 to $-$ 1.2 (corresponding to 0.2 -- 0.4 times solar;
   Vila-Costas \& Edmunds 1993; van Zee, Salzer, \& Haynes 1998). 
   Abundances in this range are typical of H~II regions in low-metallicity
   galaxies, and are interpreted in terms of the combined effects
   of primary and secondary nitrogen production (van Zee et al., and
   references therein).

   For the nucleus, FS89 have previously commented on the weakness of 
   the optical nitrogen lines relative 
   to [O~I] $\lambda$6300 and [S~II] $\lambda\lambda$6716,~6731, which
   may be taken as evidence of an underabundance of nitrogen. With the
   current dataset it is possible to estimate directly the N/O ratio within 
   the nucleus using the ratio of the O~III] $\lambda$1664/N~III] $\lambda$1750 
   lines, as discussed by Netzer (1997).  The theoretical ratio is as follows:

\begin{equation}
\frac{I(\lambda1664)}{I(\lambda1750)} 
= 0.41T_{4}^{-0.04}{\rm exp}(-0.43/T_{4}) \frac{N(O^{+2})}{N(N^{+2})},
\end{equation}
   where $T_{4}$ is the temperature in units of 10,000~K. The N$^{+2}$ and
   O$^{+2}$ are expected to show strong overlap, such that the ratio of these
   ions closely reflects the total abundance ratio.
   N~III] $\lambda$1750
   appears to be present (see Figure 1), although weak and near an artifact
   in the FOS G190H spectrum. Assuming an upper limit to the dereddened
   strength of this line to be 0.1 times that of H$\beta$, and $T =$ 15,000~K,
   equation (2) yields N(O$^{+2}$)/N(N$^{+2}$) $\geq$ 16, or log(N/O) $\leq$
   $-$1.2, consistent with the extranuclear results. Based on these
   findings, we adopt a ratio of N/O equal to 1/3 solar (i.e., N/H $=$ 1/6
   solar) for our numerical calculations.

   With the exception of N, the abundances of other elements heavier than 
   He are scaled in proportion to O. The resulting abundances, by number,
   relative to H are thus He=0.1,
   C=1.7x10$^{-4}$, O=3.4x10$^{-4}$, N=2.0x10$^{-5}$, Ne=6.0x10$^{-5}$, 
   S=8.0x10$^{-6}$, Si=1.6x10$^{-5}$, Mg=1.6x10$^{-5}$, and Fe=2.0x10$^{-5}$.
   As discussed below, gas-phase abundances are modified in some cases from 
   these values to reflect depletion onto grains.

\subsection{The Ionizing Continuum}

   In these simple models, the gas is photoionized
   by radiation from the central AGN, and therefore the results are dependent
   on what we assume for the SED and total
   luminosity of the central source. The simplest estimate of the
   shape of the SED is made by fitting the observed flux in the UV 
   to that in the soft X-ray (from {\it ROSAT}/PSPC data)
   as a simple power-law ($f_{\nu}\propto\nu^{\alpha}$).
   However, as noted by Moran et al. (1999),
   the X-ray continuum from NGC 4395 may be absorbed, and therefore
   such a fit may not represent the intrinsic SED. Using the {\it HST}/FOS 
   spectra, FHS93 fit the continuum below 2000 \AA~ with a power law
    of index 
   $=$ $-$1.
   If, for the sake of convention, we extend this to the Lyman
   limit, the continuum out to 0.1 keV 
   (the low-energy end of the {\it ROSAT} band) can be fit using a power law with an 
   index of $-$3.  This is clearly too soft to reproduce the
   observed He~II $\lambda$4686/H$\beta$ ratio. For example, a simple
   photon counting calculation, which assumes all photons between 13.6 eV and
   54.4 eV go into ionizing hydrogen while photons with energies above 54.4 eV
   go into ionizing He~II (cf. Kraemer et al. 1994), yields
   a power-law index $\sim$ $-$1.5.  
   Furthermore, Moran et al. fit the continuum between 0.1 and 2 keV with a power law 
   with a positive spectral index. This is rare in Seyfert 1 spectra, and 
   such occurrences are typically attributed to absorption by an intervening 
   layer of ``cold'' (i.e., $<$ 10$^{5}$~K) gas intrinsic to the nucleus of the
   galaxy (cf. Feldmeier et al. 1999; Kraemer et al. 1998b). Therefore, we modeled
   the continuum as a power law extending from the Lyman limit to
   1 keV, since an absorber would be relatively transparent above 1 keV, and
   Seyfert X-ray spectra often steepen below 1 keV (e.g., Arnaud et al. 1985;
   Turner \& Pounds 1989). Above 1 keV, we assumed a relatively flat
   continuum, as is typical in Seyfert galaxies (cf. Nandra
   \& Pounds 1994). The ionizing continuum is then expressed as 
   $F_{\nu}$$=$K$\nu^{\alpha}$, where

\begin{equation}
    \alpha = -1.0, ~h\nu < 13.6~{\rm eV}
\end{equation}
\begin{equation}
    \alpha = -1.7,~ 13.6eV \leq h\nu < 1000~{\rm eV}
\end{equation}
\begin{equation}
    \alpha = -0.7, ~h\nu \geq 1000~{\rm eV}.
\end{equation}

   To be conservative, we set $F_{\nu}$ at
   log($\nu$) $=$ 15.4 to the observed value from the FOS data. 
   We have estimated $F_{\nu}$ at log($\nu$) $=$ 17.4 based on a fit to the
   {\it ROSAT}/PSPC data in Moran
   et al. (1999), with the caveat that this is likely to be a lower limit if the
   spectrum is indeed absorbed at lower energies. To get the luminosity,
   we assumed a distance of 2.6 Mpc. Again, if there
   is absorption at either of these energies, the luminosity assumed here ($L_{h\nu>13.6{\rm eV}}$ $\approx$ 2.5 x 10$^{39}$
   ergs s$^{-1}$) is an underestimate, a point to which
   we will return in the Discussion section. 

\subsection{Individual Components}

  As we have noted, the spectra (see Figure 1) of NGC 4395 show emission lines from a 
  wide range ionization states (e.g. oxygen lines from all ionization
  states, from O$^{0}$ to O$^{+3}$).
  The optical and UV permitted lines possess broad (FWHM $\geq$
  1000 km s$^{-1}$) wings, which are apparently absent from even the highest
  excitation forbidden lines (e.g., [Ne~V] $\lambda$3426, 
  [O~III] $\lambda$5007).
  Therefore, a high-density component is required to fit 
  the broad emission lines. The broad H$\alpha$/H$\beta$ ratio is significantly
  greater than Case B (see Table 2), providing evidence for collisional
  excitation from the n $=$ 2 state of hydrogen, which is comparable to
  recombination at $n_{H}$ $\approx$ 10$^{10}$ cm$^{-3}$ (Krolik \& McKee 1978; 
  Ferland \& Netzer 1979). There does not appear to be a strong
  broad component of Si~III] $\lambda$1892 compared to C~III] $\lambda$1909 (although there appears to
  be some contamination by Fe~II, so it is difficult to determine this 
  accurately
  at low resolution), which constrains the upper limit for the density
  at $\sim$ 5 x 10$^{10}$ cm$^{-3}$, at which collisional de-excitation
  would strongly suppress C~III] $\lambda$1909, and, thus, the 
  emission near 1900 \AA~ would begin to be dominated by the Si~III] line,
  which has a significantly higher critical density.
  The final combination of ionization parameter and density was chosen to 
  reproduce the observed C~IV $\lambda$1550/H$\beta$ and 
  C~IV$\lambda$1550/C~III] $\lambda$1909 ratios. As such, for 
  this component (BROAD) we choose $n_{H}$ $=$ 
  3 x 10$^{10}$ cm$^{-3}$ and $U =$ 10$^{-2.25}$, at a distance $D =$ 3 x 10$^{-4}$ pc
  from the central source. For these conditions, dust temperatures 
  would be greater than the sublimation temperatures of either 
  silicate or graphite grains (i.e., in excess of 2000~K), so BROAD was
  assumed to be dust-free. The model was truncated at an 
  effective column density (the column density of ionized and
  neutral hydrogen), $N_{eff}$ $\approx$ 7 x 
  10$^{22}$~cm$^{-2}$, chosen so that the predicted broad He~II $\lambda$4686/H$\beta$
  did not fall too far below the observed value (there would be significant
  collision excitation of H$\beta$ in a warm, neutral envelope, which
  would lower this ratio).

  As noted in Section 2.1, the narrow-line spectrum shows evidence for 
  ionization stratification in the NLR. Nevertheless, we initially
  attempted to model the NLR with a single component using an average
  set of initial conditions. As expected, we found it impossible
  to produce the relative strengths of high- and low-ionization lines
  with a single set of input parameters. Consequently, multiple
  component models were required for a satisfactory fit to the emission-line ratios. 
  
  The observed [O~III] $\lambda\lambda$4959,~5007/[O~III] $\lambda$4363 is
  $\approx$ 40, which indicates either an electron temperature in excess of 
  20,000~K (Osterbrock 1974) or some modest modification of this ratio by
  collisional effects. Since it is unlikely that such a high electron
  temperature can occur in the O$^{+2}$ zone of photoionized gas, the 
  latter is more plausible. Therefore, we have modeled this component
  of the NLR (INNER) assuming a density, $n_{H}$ $=$ 10$^{6}$~cm$^{-3}$, which
  is above the critical density for the [O~III] $\lambda$5007 line. In order
  to keep from underpredicting the C~IV $\lambda$1550/C~III] $\lambda$1909
  ratio, we set $U =$ 10$^{-1.5}$,
  which places it at a distance $D =$ 0.023 pc from the central source, and
  assumed it was dust-free. Integration of the model was truncated when the
  electron temperature fell below 5000~K, and there was no longer any
  significant contribution to the line emission, making it 
  effectively radiation-bounded. 

  The [S~II] $\lambda$6716/$\lambda$6731 ratio is $\sim$ 0.9, which
  indicates an electron density $\sim$ 10$^{3}$ cm$^{-3}$; the hydrogen density
  could be somewhat higher since the ionization potential of S$^{+}$ is below
  the Lyman limit and the [S~II] lines often arise in warm, neutral
  gas. Collisional de-excitation suppresses [O~II] $\lambda$3727 
  at densities $n_{H}$ $\leq$ 10$^{4}$ cm$^{-3}$ (De Robertis \& Osterbrock
  1984), and, thus, there is a negligible contribution to the $\lambda$3727 line
  from INNER. For these reasons, we included a second NLR component (OUTER) with
  $n_{H}$ $=$ 10$^{4}$~cm$^{-3}$, $U =$ 10$^{-4}$, and $D =$ 3.76 pc.
  The weakness of narrow Mg~II $\lambda$2800/H$\beta$, particularly
  compared with other low ionization lines such as [O~I] $\lambda$6300 
  and [S~II] $\lambda\lambda$6716,6731, would indicate that there is
  likely to be depletion of magnesium onto dust grains in the lower
  ionization NLR gas. Therefore, we included cosmic dust in OUTER,
  with a dust-to-gas ratio scaled by the abundances we have chosen, i.e. approximately 50\% the 
  Galactic value. The fractional depletions are similar to those
  calculated by Seab \& Shull (1983). This component was also
  radiation-bounded.

  We attempted to fit the narrow-line ratios with a 2-component model, however
  there were several emission lines, such as [Ne~IV] $\lambda$2423 and
  [O~IV] 25.9~$\mu$m, which were underpredicted by any combination
  of the low- and high- density/ionization components described above.
  Specifically, there appears to be a component of the NLR characterized by
  low density and high ionization parameter. If this component is the 
  source of the [O~IV] 25.9~$\mu$m emission, its density must not be
  much greater than the critical density for this transition, i.e 
  about 10$^{4}$ cm$^{-3}$. Therefore, we
  added a third component (MIDDLE) with the same density and
  dust fraction as OUTER, but with $U =$ 10$^{-1.7}$, at $D =$ 0.27 pc.
  Integration was truncated at $N_{eff}$ $\approx$ 10$^{21}$cm$^{-2}$, which,
  although arbitrary, gives MIDDLE the same physical depth as OUTER.

\section{Model Results}

  The predicted line ratios for the three NLR models are listed in
  Table 3. As expected, INNER predicts strong C~IV $\lambda$1550, C~III] $\lambda$1909,
  [O~III] $\lambda$5007, and [O~III] $\lambda$4363. Nearly all the
  [O~IV] 25.9 $\mu$m and [S~IV] 10.5 $\mu$m emission comes from MIDDLE,
  which also predicts strong [O~III] $\lambda$5007, [Ne~IV] $\lambda$2324, and
  a large He~II $\lambda$4686/H$\beta$ ratio
  (due to the truncation of the partially neutral zone). Low-ionization
  lines, such as [O~II] $\lambda$3727, [S~II] $\lambda\lambda$6716,~6731,
  and [N~II] $\lambda\lambda$6548,~6583, are strongest in OUTER, except
  for Mg~II $\lambda$2800, which is affected by depletion onto dust.

  In creating a composite model from the three narrow-line components we first
  fit the low-ionization component and then added the two higher ionization
  components in a ratio to obtain the best fit to the high-ionization
  lines (such as C~IV $\lambda$1550, [Ne~IV] $\lambda$2423, and [Ne~V] 
  $\lambda$3426). The following fractional contributions to the composite
  narrow-line model were used: OUTER, 50\%; INNER, 30\%; and MIDDLE, 20\%. Although the
  fit could be marginally improved by fine-tuning the model parameters and
  the balance between these components, no significant additional
  insight would be obtained by such an exercise. This three-component
  model is, of course, a very simple approximation of what is likely to be a complex region
  consisting of a range of physical conditions. Nevertheless, we can derive 
  some insight into the global physical parameters from these models.
  
  The comparison of the composite model predictions and the dereddened
  narrow-line ratios is given in Table 4. Overall, the fit is quite good
  for lines from a variety of ionization states and critical densities,
  such as Si~IV $\lambda$1398 $+$ O~IV] $\lambda$1402, C~IV $\lambda$1550, 
  [Ne~IV] $\lambda$2423, [O~II] $\lambda$3727, [O~I] $\lambda\lambda$6300,~6364,
  [S~II] $\lambda\lambda$6716,~6731, and
  many weaker lines. This indicates that the models provide a 
  reasonable representation of the range in ionization parameter and 
  density of the NLR gas. Also, the quality of the fit for the nitrogen
  lines is evidence that the chosen N/O ratio is approximately correct. The 
  predicted strengths of the coronal lines, such as [Fe~VII] $\lambda$6087 and 
  [Ne~V] $\lambda\lambda$3346,~3426, are a reasonable match to the 
  observations; hence, there is 
  no need to include the additional collisional heating (due to shocks) suggested by
  Contini (1997).

  There are, as is typical of such simple models, a few discrepancies.
  First, the predicted He~II lines are a bit strong. Although this may 
  be due to the arbitrary truncation of MIDDLE, it may also be an
  indication that the SED we have assumed for the models is a bit too
  hard. As noted in Section 3.2, in fitting the UV to X-ray continuum, we have 
  not corrected the UV flux to account for reddening. If the 
  reddening is not negligible, and the
  derived X-ray flux at 1 keV is not biased by the strong absorption 
  apparent at $\sim$ 0.1 keV, the intrinsic continuum may be somewhat
  softer. However, the spectral index derived from photon counting is
  close to the value we used ($-$1.5, as opposed to $-$1.7) and, therefore, we do not think that
  the intrinsic SED differs significantly from that which we have assumed.
  The [O~III] $\lambda$5007/[O~III] $\lambda$4363 ratio
  is a bit low, which could be remedied if the density for INNER were
  decreased by a factor $\sim$ 2. This would also help reduce the relative
  strength of the C~III] $\lambda$1909 line, which is enhanced at the
  higher density.  Mg~II $\lambda$2800 is 
  a bit too strong, indicating that some dust may exist in the
  inner NLR, which is plausible, since dust temperatures calculated for
  INNER would not exceed 700~K. 

  Although these discrepancies could be eliminated with minor adjustments
  of free parameters, the underpredictions of the strengths of 
  [O~IV] 25.9 $\mu$m and [S~IV] 10.5 $\mu$m are
  not as easily remedied. Since the strengths of lines from the same 
  ionization states, such as O~IV] $\lambda$1402, are not similarly
  underpredicted, and the conditions in MIDDLE have been optimized for
  the high-ionization IR fine structure lines, the simplest
  explanation is that these lines arise in gas that
  is obscured by a layer of dust, and therefore not detected in the
  UV or optical. If the conditions are identical to those in MIDDLE
  ([O~IV] 25.9~$\mu$m/H$\beta$ $=$ 2.0), the 
  contribution from the obscured component would have to be a factor
  of $\sim$ 4 times greater than MIDDLE, and its contribution
  to the observed narrow 
  H$\beta$ must be negligible (i.e., $<$ 10\%). Therefore, the ratio of 
  ([O~IV] 25.9~$\mu$m/H$\beta$)$_{observed}$/([O~IV] 25.9~$\mu$m/H$\beta$)$_{emitted}$
  would be $\sim$ 10. From this ratio, we can derive an estimate of the
  column density of the obscuring layer. Assuming no extinction at 25.9~$\mu$m, and the reddening
  curve of Savage \& Mathis (1979), the reddening derived from the line ratios
  is $E_{B-V}$ $\approx$ 0.68 mag. This
  yields a column density $N_{eff}$ $\approx$ 7 x 10$^{21}$ cm$^{-2}$, using
  the relation given in Shull \& Van Steenberg (1985), scaled by the
  abundances we have assumed for NGC 4395. Although this is somewhat larger
  than the $N_{eff}$ for either of the low-ionization models, the
  truncation of each was arbitrary, and a larger column for
  OUTER would not appreciably affect the predicted emission-line ratios. 
  Thus, it is probable that the [O~IV] emission arises in a region
  that is obscured by a layer of dusty emission line gas. If so, there
  are consequences for the estimates of the covering factor of the
  ionized gas and the intrinsic luminosity of the central source, which we
  discuss in Section 5.

  The comparison of the predictions from BROAD and the
  dereddened broad-line ratios is given in Table 5. Again, the overall 
  agreement is quite good, in particular the ratio of C~IV $\lambda$1550/C~III]
  $\lambda$1909, and the strengths of the Si~IV $\lambda$1398 $+$ 
  O~IV] $\lambda$1402 and the He~II/H$\beta$ ratios.
  This is interesting, since we chose to represent the BLR with
  a single (i.e., average) set of physical conditions, while it is apparent that the region
  is extended, as evidenced by the fact that C~IV $\lambda$1550 has
  broader wings than H$\beta$. The single discrepancy is the overprediction of the
  Mg~II $\lambda$2800 line. It is unlikely that the weakness of the
  observed line is due to depletion of magnesium onto dust grains, since, as noted above,
  dust would probably be unable to exist in the BLR of NGC 4395. It is
  possible that the relative abundances of magnesium and other refractory
  elements, such as silicon and iron, are somewhat less than what we have assumed for these models,
  although there is no strong evidence for this, based on the narrow emission
  line spectrum. Since much of the
  Mg~II $\lambda$2800 emission occurs at optical depths $>$ 5 at the
  Lyman limit, it is possible that the physical size of the broad-line
  clouds is smaller than what we have assumed. If so, a model characterized
  by a lower value for $U$, truncated at a smaller $N_{eff}$, might result
  in a slightly better fit.

\section{Discussion}

  As noted above, these models provide a very good fit to the observed
  emission-line ratios. Also, the radial distance of OUTER, the 
  component furthest from the central source, is 3.76 pc, which is well within
  the upper limit of 13 pc for the FWHM of the emission-line region (FHS93).
  Therefore, we feel confident that the model results can be used to
  estimate some of the global properties of the active nucleus of NGC 4395.

  The emitted H$\beta$ flux predicted by each NLR model is given in the footnotes
  for Table 3. From these values, the luminosity assumed for the ionizing
  source, and the fractional contribution to the
  narrow H$\beta$ luminosity of NGC 4395, we may estimate the ``covering
  factor'' of each component, or more specifically, the fraction of
  ionizing photons intercepted by each component and converted into
  emission-line photons. If, for the sake of simplicity, we assume that
  each component can be approximated as a fraction of the surface of 
  a sphere with a radius equal to the distance from the central
  source, we derive the following covering factors: INNER, 27\%;
  MIDDLE, 61\%, and OUTER, 42\%. The emitted H$\beta$ flux for BROAD
  is given in Table 5. By comparison to the observed broad H$\beta$ we
  calculate a covering factor of 67\% for BROAD. Thus, the total covering factor
  of the emission-line gas is $\sim$ 200\%.  This implies that
  the luminosity calculated from the observed continuum fluxes
  is an underestimate of at least a factor of 2. 
  However, as discussed above, there
  is also evidence for a significant amount of obscured gas not included in our
  models. If we include this additional component,
  the total covering would increase by a factor greater than 2.

  As noted in Section 3.2, there is evidence that the observed UV to
  X-ray continuum may be absorbed. Since the covering factor of the 
  emission-line gas is large, it is possible that 
  the continuum source is occulted by it. In Figure 4, we compare the
  incident and transmitted ionizing continnua for model MIDDLE,
  the NLR component with the largest covering factor.
  Since MIDDLE is essentially opaque at the He~II Lyman limit, a fit 
  to the 0.1 - 2 keV band yields a positive index, as seen in the 
  {\it ROSAT}/PSPC spectrum (Moran et al. 1999). There is also
  attenuation below the Lyman limit due to dust. The amount of reddening
  in MIDDLE is $E_{B-V}$ $\approx$ 0.10 mag, which is only
  slightly larger than the reddening determined from the emission lines,
  $E_{B-V}$ $=$ 0.05 $\pm$ 0.03.  If the continuum is reddened by this
  amount, the intrinsic luminosity in the UV could easily be
  more than a factor of 2 greater than what we have assumed, based on the 
  observed fluxes. In fact, depending on the strength of the 2200 \AA~ 
  bump intrinsic to NGC 4395, the continuum may indeed be somewhat more
  reddened than the emission lines (Moran et al. 1999). We are led to the 
  following conclusion: the covering factor of the
  emission line gas is near unity and the total luminosity of the central
  source in the Lyman continuum may be more than 4 times the estimate 
  based on the observed fluxes, or 
  $L_{h\nu>13.6{\rm eV}}$ $\sim$ 1 x 10$^{40}$ ergs s$^{-1}$.
  If so, the radial distances of each of the model components would
  increase by more than a factor of 2. 
  
  The observed equivalent width (EW) of the broad C~IV $\lambda$1550,
  76~\AA~ (FHS93), is a factor of 10 lower than predicted by an 
  extrapolation of the Baldwin relation, a negative correlation between
  the EW and source luminosity (see Figure 3(a) in 
  Kinney, Rivolo \& Koratkar (1990)). However, as the covering factor
  of the emission-line gas approaches unity, the EW must approach a maximum,
  since at this point the conversion of continuum photons into 
  line photons must saturate. This must be the case for NGC 4395,
  since the covering factor of the BLR gas is large. If NGC 4395 were
  included in the plot in Kinney et al. a turn-over at the 
  low-luminosity end would be apparent. Evidence that the Baldwin
  effect flattens in slope at low luminosity has been reported
  previously by several researchers (e.g., V\'eron-Cetty, V\'eron, 
  \& Tarenghi 1983;
  Wu, Boggess, \& Gull 1983; Kinney et al. 1990; Osmer, Porter, \& Green 1994). The high
  covering factor determined for NGC 4395 lends support to the
  idea that the Baldwin effect is at least partially driven by a 
  luminosity dependence in covering factor in AGN; the curvature seen in
  the Baldwin relation can then be naturally explained by coverage
  hitting a maximum value at low luminosities (Wampler et al. 1984). 

  We can also use the model results to determine the mass of the putative
  central black hole. As noted in Section 2.1, the FWHM of the broad 
  H$\beta$ line is $\sim$
  1500 km s$^{-1}$. From the radial distance assumed for BROAD,
  3 x 10$^{-4}$ pc, we compute a virial mass, $M_{bh}$ $\sim$ 1.5 x 10$^{5}$
  $M_{\odot}$. which is similar to the mass derived from the 
  stellar kinematics (Filippenko \& Ho 1999), although if we scale the radial distance 
  of BROAD by the ratio of the
  intrinsic to observed central source luminosity, $M_{bh}$ would be
  twice as large. Interestingly, the width of the broad C~IV $\lambda$1550 line 
  ($\sim$ 5000 km s$^{-1}$) implies a mass an order of magnitude
  larger, although the BLR of NGC 4395 is stratified and
  it is likely that much of the C~IV emission arises closer to the 
  nucleus than BROAD. Also, non-gravitational effects, such as a wind,
  could result in a radial component of the gas motion that could
  bias the derivation of the central mass. As we also noted in Section 2.1, the observed widths of the narrow lines are quite 
  small, with FWHM $\leq$ 50 km s$^{-1}$ for [O~III] $\lambda$5007
  (FS89). From our models, the average distance of the [O~III] emitting
  clouds is 0.14 pc. Based on our mass estimate, the velocity width for 
  clouds in virial motion would be $\sim$ 55 km s$^{-1}$, in 
  agreement with the observations.

  The model results support our assumptions regarding the elemental abundances 
  in the nucleus of NGC 4395. First, the heavy element abundances are clearly subsolar, as
  previous studies of the H~II regions have also indicated (Roy et al. 1996). 
  Although most AGN appear to possess solar or supersolar abundances (cf. Ferland et 
  al. 1996; Oliva 1996; Netzer \& Turner 1997), NGC 4395 is an extremely 
  low-mass, low-luminosity AGN, and therefore the underabundance
  of heavy elements is to be expected. The underabundance of nitrogen with
  respect to oxygen is also typical of low metallicity dwarf galaxies
  (cf. van Zee et al. 1998), and is likely to be the result of 
  the combined effects of primary and secondary origins of 
  nitrogen.

\section{Conclusions}

  We have analyzed UV, optical, and IR spectra of the active nucleus in the Sd IV
  galaxy NGC 4395. The permitted lines in the UV and optical were
  deconvolved into broad (FWHM $>$ 1000 km s$^{-1}$) and narrow
  components. We have constructed photoionization models of the
  broad-line and narrow-line gas, and have successfully
  matched the observed emission-lines ratios, with the very few
  exceptions noted in Section 4. The model results predict
  an NLR size ($<$ 5 pc in radius) that is within the observed constraints.
  The models predict a covering factor for the emission-line gas
  greater than unity, but we have shown evidence that the observed
  continuum in the UV and X-ray is absorbed by an intervening layer
  of dusty gas which has properties similar to the low-density components
  used in our composite NLR model. Our analysis supports
  previous conclusions regarding the nature of NGC 4395 (FS89; FHS93),
  specifically that this object harbors a dwarf Seyfert 1 nucleus.

  Our models predict that the NLR in NGC 4395 is extremely compact. In particular, the 
  region in which most of the [O~III] $\lambda$5007 emission arises is 
  $<$ 1/2 pc from the central source. Also, the recombination
  time for INNER is roughly a few months. As such, if there were changes
  in the continuum flux on similar timescales (a few months to several years),
  we might expect to see 
  corresponding variations in the narrow emission lines. Although the optical 
  continuum flux from 
  NGC 4395 may vary as much as 20\% on timescales $\sim$ 1 day 
  (Lira, Lawrence, \& Johnson 1999), the nature of the variability over longer
  timescales in not well known. As such, continued monitoring of NGC 4395 to
  look for both long-term continuum changes and possible narrow-line 
  variability is warranted.
  
  Perhaps the most important conclusion is, simply, that NGC 4395 is an
  example of the AGN phenomenon extended to a low luminosity extreme
  ($\sim$ 10$^{6}$ fainter than typical QSOs). It is also the only known
  example of an active nucleus within a bulge-less,
  extreme late-type galaxy. It is fascinating that
  the fundamental physics of an AGN, specifically a region powered by
  a massive black hole, is at work over such a huge range in mass
  and luminosity. In fact, it is worth noting that it is likely that
  an AGN this weak could only be detected in a low-mass galaxy,
  since any other nuclear activity, such as a starburst, could easily
  overpower it. Being able to study the low-luminosity end of the AGN 
  menagerie also gives us insight into properties that appear to be
  a function of luminosity, such as the Baldwin effect. These results
  help amplify the importance of covering factor for this effect, and are further evidence that large 
  BLR covering factors may be a property of low-luminosity AGNs.

\acknowledgments

 S.B.K is grateful to Jane Turner for helpful discussions concerning the
 {\it ROSAT}/PSPC data. J.C.S. thanks 
 Sarah Unger and the rest of the IPAC staff for their assistance in 
 scheduling and analyzing observations with {\it ISO}.
 We acknowledge support from the following NASA grants: NAG 5-4103 (S.B.K. 
 and D.M.C.), AR-07527 (L.C.H. and A.V.F.), NAG 5-3563 (J.C.S.), and 
 NAG 5-3556 (A.V.F.).

\clearpage

\figcaption[Kraemer_fig1.eps]{Combined FOS (G130H, G190H, G270H) and 
ground-based spectra of the nucleus of NGC 4395, from 1400 to 10000 \AA~.
}\label{fig1}

\figcaption[Kraemer_fig2.ps]{Profiles of C~IV $\lambda$1550 ({\it top})
and H$\beta$ ({\it bottom}). Note the broad wings and narrow cores of
each line. The widths of the narrow cores are determined by the
instrumental resolution (see text). The features at 1544~\AA~
and 1550~\AA~ are due to absorption, possibly intrinsic to NGC 4395.
}\label{fig2}

\figcaption[Kraemer_fig3.ps]{{\it ISO} spectra of NGC 4395. The vertical dotted
lines indicate the wavelengths of the emission lines ([Si~X] 3.9357 $\mu$m, 
[S~IV] 10.5105 $\mu$m, [S~III] 33.4810 $\mu$m, and [O~IV] 25.8903 $\mu$m) in the 
four spectra.
}\label{fig3}

\figcaption[Kraemer_fig4.eps]{Comparison of the incident UV to X-ray continuum 
({\it solid line}) and that transmitted through component MIDDLE ({\it dotted 
line}). 
Note the deep absorption edges at the Lyman limit (13.6 eV) and 
He~II Lyman limit (54.4 eV), and weaker edges from He~I (24.6 eV) and 
C~III (47.4 eV).
}\label{fig4}

\clearpage
\begin{deluxetable}{lrr}
\tablecolumns{3}
\footnotesize
\tablecaption{Observed Narrow Emission-Line Ratios for NGC 4395 
(relative to H$\beta$$^{a}$)\label{tbl-1}}
\tablewidth{0pt}
\tablehead{
\colhead{Line} & \colhead{~Observed~} & \colhead{~Dereddened$^{b}$~}
}
\startdata
Si IV $\lambda$1398$+$ O IV] $\lambda$1402  &0.58 ($\pm$0.09) &0.72 ($\pm$0.14) \\
N IV] $\lambda$1486         	      	  &0.09 ($\pm$0.03) &0.11 ($\pm$0.04) \\
C IV $\lambda$1550          	      	  &2.49 ($\pm$0.34) &3.04 ($\pm$0.52) \\
He II $\lambda$1640         	      	  &0.75 ($\pm$0.11) &0.91 ($\pm$0.15) \\
O III] $\lambda$1663        	      	  &0.39 ($\pm$0.07) &0.47 ($\pm$0.09) \\
Ne III] $\lambda$1815                     &0.03 ($\pm$0.01) &0.04 ($\pm$0.01) \\
Si III] $\lambda$1882,1892	          &0.15 ($\pm$0.04) &0.18 ($\pm$0.05) \\
C III] $\lambda$1909                      &0.83 ($\pm$0.11) &1.02 ($\pm$0.17) \\
C II] $\lambda$2326                       &0.38 ($\pm$0.06) &0.48 ($\pm$0.09) \\
$[$Ne IV] $\lambda$2423     	      	  &0.23 ($\pm$0.05) &0.28 ($\pm$0.06) \\
$[$O II] $\lambda$2470      	      	  &0.11 ($\pm$0.03) &0.13 ($\pm$0.04) \\
Mg II $\lambda$2800         	      	  &0.25 ($\pm$0.05) &0.28 ($\pm$0.05) \\
O III $\lambda$3133			  &0.22 ($\pm$0.03) &0.24 ($\pm$0.05) \\
He II $\lambda$3204                   	  &0.09 ($\pm$0.02) &0.10 ($\pm$0.02) \\
$[$Ne V] $\lambda$3346      	      	  &0.13 ($\pm$0.03) &0.14 ($\pm$0.03) \\
$[$Ne V] $\lambda$3426      	      	  &0.26 ($\pm$0.04) &0.28 ($\pm$0.04) \\
$[$O II] $\lambda$3727      	      	  &1.73 ($\pm$0.17) &1.81 ($\pm$0.17) \\
$[$Fe VII] $\lambda$3760    	      	  &0.04 ($\pm$0.01) &0.04 ($\pm$0.01) \\
H9$\lambda$3835                           &0.02 ($\pm$0.01) &0.02 ($\pm$0.01) \\
$[$Ne III] $\lambda$3869    	      	  &0.90 ($\pm$0.09) &0.93 ($\pm$0.09) \\
H8, He I $\lambda$3889                    &0.14 ($\pm$0.02) &0.15 ($\pm$0.02) \\
$[$Ne III] $\lambda$3968, H$\epsilon$     &0.41 ($\pm$0.04) &0.42 ($\pm$0.04) \\
$[$S II] $\lambda$4072                	  &0.24 ($\pm$0.04) &0.25 ($\pm$0.04) \\
H$\delta$$\lambda$4102 	      	          &0.21 ($\pm$0.04) &0.22 ($\pm$0.04) \\
H$\gamma$ $\lambda$4340      	      	  &0.37 ($\pm$0.05) &0.38 ($\pm$0.05) \\
$[$O III] $\lambda$4363      	      	  &0.26 ($\pm$0.03) &0.27 ($\pm$0.03) \\
He I $\lambda$4471                        &0.03 ($\pm$0.01) &0.03 ($\pm$0.01) \\
Fe III $\lambda$4661			  &0.05 ($\pm$0.01) &0.05 ($\pm$0.01) \\
He II $\lambda$4686          	      	  &0.12 ($\pm$0.02) &0.12 ($\pm$0.02) \\
\tablebreak
$[$Ar IV]$\lambda$4711~~~~~~~~~~~~~~~~~                   &0.04 ($\pm$0.01) &0.04 ($\pm$0.01) \\
$[$Ar IV]$\lambda$4740                    &0.05 ($\pm$0.01) &0.05 ($\pm$0.01) \\
H$\beta$                 	      	  &1.00~~~~~~~~~~~  &1.00~~~~~~~~~~~ \\
$[$O III] $\lambda$4959                   &2.74 ($\pm$0.27) &2.70 ($\pm$0.27) \\
$[$O III] $\lambda$5007                   &8.32 ($\pm$0.79) &8.28 ($\pm$0.79) \\
$[$N I] $\lambda$5200                    &0.07 ($\pm$0.01) &0.07 ($\pm$0.01) \\
He II $\lambda$5412                       &0.02 ($\pm$0.01) &0.02 ($\pm$0.01) \\
$[$Fe VII] $\lambda$5721     	      	  &0.03 ($\pm$0.01) &0.03 ($\pm$0.01) \\
$[$N II] $\lambda$5755                    &0.02 ($\pm$0.01) &0.02 ($\pm$0.01) \\
He I $\lambda$5876           	      	  &0.13 ($\pm$0.02) &0.13 ($\pm$0.01) \\
$[$Fe VII] $\lambda$6087     	      	  &0.03 ($\pm$0.01) &0.03 ($\pm$0.01) \\
$[$O I] $\lambda$6300        	      	  &1.05 ($\pm$0.10) &1.00 ($\pm$0.10) \\
$[$O I] $\lambda$6364                     &0.34 ($\pm$0.04) &0.32 ($\pm$0.04) \\
$[$N II] $\lambda$6548                	  &0.25 ($\pm$0.08) &0.24 ($\pm$0.08) \\
H$\alpha$ $\lambda$6563      	      	  &3.24 ($\pm$0.38) &3.07 ($\pm$0.39) \\
$[$N II] $\lambda$6583                	  &0.75 ($\pm$0.14) &0.71 ($\pm$0.14) \\
He I$\lambda$6680                         &0.04 ($\pm$0.01) &0.04 ($\pm$0.01) \\
$[$S II] $\lambda$6716	                  &0.91 ($\pm$0.08) &0.86 ($\pm$0.09) \\
$[$S II] $\lambda$6731	                  &1.04 ($\pm$0.10) &0.98 ($\pm$0.10) \\
He I $\lambda$7065                        &0.08 ($\pm$0.02) &0.07 ($\pm$0.02) \\
$[$Ar III] $\lambda$7136		  &0.24 ($\pm$0.03) &0.22 ($\pm$0.03) \\
$[$O II] $\lambda$7325			  &0.37 ($\pm$0.05) &0.35 ($\pm$0.05) \\
$[$Ar III] $\lambda$7751		  &0.07 ($\pm$0.02) &0.06 ($\pm$0.02) \\
O I $\lambda$8446			  &0.09 ($\pm$0.02) &0.08 ($\pm$0.02) \\
$[$S III] $\lambda$9069 		  &0.53 ($\pm$0.09) &0.48 ($\pm$0.09) \\
$[$S III] $\lambda$9532                   &2.22 ($\pm$0.25) &2.00 ($\pm$0.28) \\
\tablebreak
$[$Si IX] 3.9357 $\mu$m$^{c}$~~~~~~~~~~~~~~                     &$<$0.07~~~~~~~~ 
 &$<$0.06~~~~~~~~ \\
$[$S IV] 10.5105 $\mu$m                    &0.91 ($\pm$0.33)  &0.76 ($\pm$0.28)\\
$[$O IV] 25.8903 $\mu$m                    &2.28 ($\pm$0.38)  &1.90 ($\pm$0.38)\\
$[$S III] 33.4810  $\mu$m$^{c}$                   &$<$1.49~~~~~~~~     &$<$1.24~~~~~~~~\\

\tablenotetext{a}{Flux (narrow H$\beta$) $=$ (2.57 $\pm$0.24) x 10$^{-14}$
ergs s$^{-1}$ cm$^{-2}$.}
\tablenotetext{b}{E$_{B-V}$ $=$ 0.05 $\pm$0.03 mag.}
\tablenotetext{c}{3$\sigma$ upper limit.}
\enddata
\end{deluxetable}
\clearpage
\begin{deluxetable}{lrr}
\tablecolumns{3}
\footnotesize
\tablecaption{Broad-Line Ratios for NGC 4395 
(relative to broad H$\beta$$^{a}$)\label{tbl-2}}
\tablewidth{0pt}
\tablehead{
\colhead{Line} & \colhead{~Observed~} 
& \colhead{~Dereddened$^{b}$~}
}
\startdata
Si IV $\lambda$1398 $+$ O IV] $\lambda$1402 &0.64 ($\pm$0.25) &0.80 ($\pm$0.35)\\
C IV $\lambda$1550          	      	 &4.00 ($\pm$0.74) &4.89 ($\pm$1.06)\\
He II $\lambda$1640         	      	 &0.92 ($\pm$0.31) &1.12 ($\pm$0.40)\\
C III] $\lambda$1909                     &1.20 ($\pm$0.27) &1.47 ($\pm$0.37)\\
Mg II $\lambda$2800                      &0.33 ($\pm$0.12) &0.37 ($\pm$0.14)\\
H$\delta$ $\lambda$4102                  &0.20 ($\pm$0.07) &0.21 ($\pm$0.07)\\
H$\gamma$ $\lambda$4340                  &0.42 ($\pm$0.12) &0.43 ($\pm$0.12)\\
He II $\lambda$4686          	      	 &0.19 ($\pm$0.10) &0.19 ($\pm$0.10)\\
He I $\lambda$5876           	      	 &0.13 ($\pm$0.04) &0.13 ($\pm$0.04)\\
H$\alpha$ $\lambda$6563      	      	 &5.42 ($\pm$0.93) &5.13 ($\pm$0.90)\\
\tablenotetext{a}{Flux (broad H$\beta$) $=$ (2.07 $\pm$0.31) x 10$^{-14}$
ergs s$^{-1}$ cm$^{-2}$.}
\tablenotetext{b}{E$_{B-V}$ $=$ 0.05 $\pm$0.03 mag.}
\enddata
\end{deluxetable}

\clearpage
\begin{deluxetable}{lrrr}
\tablecolumns{4}
\footnotesize
\tablecaption{Model$^{a}$ Predictions for Narrow-Line Ratios
(relative to H$\beta$)\label{tbl-3}}
\tablewidth{0pt}
\tablehead{
\colhead{Line} & \colhead{~INNER$^{b}$~} 
& \colhead{~MIDDLE$^{c}$~}
& \colhead{~OUTER$^{d}$~}
}
\startdata
Ly$\alpha$   &40.09 & 6.63 &36.51 \\
Si IV $\lambda$1398        & 0.30  & 0.03 &0.00 \\
O IV] $\lambda$1402  & 0.84 & 1.91 & 0.00 \\
N IV] $\lambda$1486       & 0.22  & 0.58 & 0.00\\
C IV $\lambda$1550       & 9.79 & 2.00 & 0.00 \\
He II $\lambda$1640         & 1.14 & 3.32 & 0.55\\
O III] $\lambda$1663        	  & 1.12 & 1.51 & 0.01\\
N III] $\lambda$1750  & 0.14 & 0.30 & 0.00\\
Ne III] $\lambda$1815            & 0.01 & 0.01 & 0.00\\
Si III] $\lambda$1882	      & 0.00 & 0.02 & 0.00\\
Si III] $\lambda$1892	       & 0.30 & 0.03 & 0.00\\
C III] $\lambda$1909         & 4.20 & 2.54 & 0.07\\
$[$O III] $\lambda$2321    & 0.31 & 0.15 & 0.00 \\
C II] $\lambda$2326              & 0.28 & 0.03 & 0.51\\
$[$Ne IV] $\lambda$2423     	 & 0.05 & 1.30 & 0.00\\
$[$O II] $\lambda$2470      	 & 0.06 & 0.02 & 0.49\\
Mg II $\lambda$2800         	 & 0.82 & 0.01 & 0.38\\
He II $\lambda$3204              & 0.07 & 0.20 & 0.03 \\
$[$Ne V] $\lambda$3346      	 & 0.17 & 0.52 & 0.00\\
$[$Ne V] $\lambda$3426      	 & 0.45 & 1.42 & 0.00 \\
$[$O II] $\lambda$3727      	 & 0.02 & 0.09 & 4.23 \\
$[$Fe VII] $\lambda$3760    	 & 0.05 & 0.03 & 0.00 \\
$[$Ne III] $\lambda$3869    	 & 1.71 & 1.17 & 0.90 \\
$[$Ne III] $\lambda$3968      & 0.53 & 0.36 & 0.28 \\
$[$S II] $\lambda$4072             & 0.14 & 0.00 & 0.53\\
H$\delta$$\lambda$4102 	      	    & 0.26 & 0.26 & 0.26\\
H$\gamma$ $\lambda$4340      	    & 0.47 & 0.47 & 0.47\\
$[$O III] $\lambda$4363      	    & 0.89 & 0.52 & 0.00\\
\tablebreak
He I $\lambda$4471                   & 0.04 & 0.03 & 0.05\\
$[$Mg I] $\lambda$4571               & 0.00 & 0.00 & 0.12\\
He II $\lambda$4686          	     & 0.16 & 0.48 & 0.08\\
H$\beta$                 	    & 1.00  & 1.00 & 1.00\\
$[$O III] $\lambda$4959             & 3.13 & 6.62 & 0.14\\
$[$O III] $\lambda$5007             & 9.40 &19.88 & 0.43 \\
$[$N I] $\lambda$5200               & 0.01 & 0.00 & 0.24 \\
He II $\lambda$5412                 & 0.01 & 0.04 & 0.01 \\
$[$Fe VII] $\lambda$5721     	    & 0.06 & 0.03 & 0.00\\
$[$N II] $\lambda$5755              & 0.01 & 0.00 & 0.03\\
He I $\lambda$5876           	    & 0.10 & 0.07 & 0.15\\
$[$Fe VII] $\lambda$6087     	    & 0.09 & 0.04 & 0.00\\
$[$O I] $\lambda$6300        	    & 0.55 & 0.00 & 1.50\\
$[$O I] $\lambda$6364               & 0.18 & 0.00 & 0.48\\
$[$N II] $\lambda$6548              & 0.01 & 0.01 & 0.42\\
H$\alpha$ $\lambda$6563      	    & 2.86 & 2.90 & 3.01\\
$[$N II] $\lambda$6583              & 0.02 & 0.01 & 1.23\\
$[$S II] $\lambda$6716              & 0.06 & 0.00 & 1.56\\
$[$S II] $\lambda$6731              & 0.13 & 0.00 & 1.91\\
$[$O II] $\lambda$7325	            & 0.08 & 0.02 & 0.62\\
$[$S III] $\lambda$9069             & 0.18 & 0.12 & 0.68 \\
\tablebreak
$[$S III] $\lambda$9532            & 0.43 & 0.30 & 1.66\\
$[$S IV] 10.5105 $\mu$m             & 0.04 & 1.39 & 0.00\\
$[$Ne III] 15.3843 $\mu$m           & 0.35 & 0.86 & 1.96\\ 
$[$O IV] 25.8903 $\mu$m             & 0.01 & 2.03 & 0.00\\
$[$S III] 33.4810 $\mu$m            & 0.01 & 0.01 & 0.59\\
\tablenotetext{a}{1/2 solar abundances, with 1/6 solar N/H.}
\tablenotetext{b}{U $=$ 10$^{-1.5}$, n$_{H}$ $=$ 1 x 10$^{6}$ cm$^{-3}$, 
no dust, radiation bounded, flux (H$\beta$ at ionized face) 
$=$ 3.64 x 10$^{2}$ ergs s$^{-1}$ cm$^{-2}$.}
\tablenotetext{c}{U $=$ 10$^{-1.7}$, n$_{H}$ $=$ 1 x 10$^{4}$ cm$^{-3}$, dust 
included, N$_{eff}$ $\sim$ 10$^{21}$
cm$^{-2}$, flux (H$\beta$ at ionized face) 
$=$ 8.09 x 10$^{-1}$ ergs s$^{-1}$ cm$^{-2}$.}
\tablenotetext{d}{U $=$ 10$^{-4.0}$, n$_{H}$ $=$ 1 x 10$^{4}$ cm$^{-3}$, dust 
included, radiation bounded, flux (H$\beta$ at ionized face) 
$=$ 1.47 x 10$^{-2}$ ergs s$^{-1}$ cm$^{-2}$.}
\enddata
\end{deluxetable}

\clearpage
\begin{deluxetable}{lrr}
\tablecolumns{3}
\footnotesize
\tablecaption{Comparison of Narrow-Line Ratios 
(relative to H$\beta$$^{a}$) to Composite NLR Model Predictions\label{tbl-4}}
\tablewidth{0pt}
\tablehead{
\colhead{Line} & \colhead{~Dereddened$^{b}$~}
& \colhead{~Composite Model$^{c}$}
}
\startdata
Si IV $\lambda$1398 $+$ O IV] $\lambda$1402 &0.72 ($\pm$0.14) &0.73\\
N IV] $\lambda$1486         	      	 &0.11 ($\pm$0.04) &0.18\\
C IV $\lambda$1550          	      	  &3.04 ($\pm$0.52) &3.33\\
He II $\lambda$1640         	      	  &0.91 ($\pm$0.15) &1.28\\
O III] $\lambda$1663        	      	  &0.47 ($\pm$0.09) &0.64\\
Ne III] $\lambda$1815                     &0.04 ($\pm$0.01) &0.01\\
Si III] $\lambda$1882,1892	          &0.18 ($\pm$0.05) &0.10\\
C III] $\lambda$1909                      &1.02 ($\pm$0.17) &1.80\\
C II] $\lambda$2326                       &0.48 ($\pm$0.09) &0.34\\
$[$Ne IV] $\lambda$2423     	      	  &0.28 ($\pm$0.06) &0.28\\
$[$O II] $\lambda$2470      	      	  &0.13 ($\pm$0.04) &0.27\\
Mg II $\lambda$2800         	      	  &0.28 ($\pm$0.05) &0.44\\
O III $\lambda$3133			  &0.24 ($\pm$0.05) &-\\
He II $\lambda$3204                   	  &0.10 ($\pm$0.02) &0.08\\
$[$Ne V] $\lambda$3346      	      	  &0.14 ($\pm$0.03) &0.15\\
$[$Ne V] $\lambda$3426      	          &0.28 ($\pm$0.04) &0.42\\
$[$O II] $\lambda$3727      	      	  &1.81 ($\pm$0.17) &2.14\\
$[$Fe VII] $\lambda$3760    	      	  &0.04 ($\pm$0.01) &0.02\\
H9$\lambda$3835                           &0.02 ($\pm$0.01) &-\\
$[$Ne III] $\lambda$3869    	      	  &0.93 ($\pm$0.09) &1.20\\
H8, He I $\lambda$3889                    &0.15 ($\pm$0.02) &-\\
$[$Ne III] $\lambda$3968, H$\epsilon$     &0.42 ($\pm$0.04) &0.37 (w/o H$\epsilon$)\\
$[$S II] $\lambda$4072                	  &0.25 ($\pm$0.04) &0.31\\
H$\delta$$\lambda$4102 	      	          &0.22 ($\pm$0.04) &0.26\\
H$\gamma$ $\lambda$4340      	      	  &0.38 ($\pm$0.05) &0.47\\
$[$O III] $\lambda$4363      	      	  &0.27 ($\pm$0.03) &0.37\\
He I $\lambda$4471                        &0.03 ($\pm$0.01) &0.04\\
Fe III $\lambda$4661 			  &0.05 ($\pm$0.01) &-\\
He II $\lambda$4686          	      	  &0.12 ($\pm$0.02) &0.18\\
\tablebreak
$[$Ar IV]$\lambda$4711~~~~~~~~~~~~~~~~~   &0.04 ($\pm$0.01) &-\\
$[$Ar IV]$\lambda$4740                    &0.05 ($\pm$0.01) &-\\
H$\beta$                 	      	  &1.00~~~~~~~~~~~  &1.00  \\
$[$O III] $\lambda$4959                   &2.70 ($\pm$0.27) &2.34\\
$[$O III] $\lambda$5007                   &8.28 ($\pm$0.79) &7.01\\
$[$N I] $\lambda$5200                    &0.07 ($\pm$0.01) &0.12\\
He II $\lambda$5412                       &0.02 ($\pm$0.01) &0.01\\
$[$Fe VII] $\lambda$5721     	      	  &0.03 ($\pm$0.01) &0.02\\
$[$N II] $\lambda$5755                    &0.02 ($\pm$0.01) &0.02\\
He I $\lambda$5876           	      	  &0.13 ($\pm$0.01) &0.12\\
$[$Fe VII] $\lambda$6087     	          &0.03 ($\pm$0.01) &0.04\\
$[$O I] $\lambda$6300        	      	  &1.00 ($\pm$0.10) &0.91\\
$[$O I] $\lambda$6364                     &0.32 ($\pm$0.04) &0.29\\
$[$N II] $\lambda$6548                	  &0.24 ($\pm$0.08) &0.22\\
H$\alpha$ $\lambda$6563      	      	  &3.07 ($\pm$0.39) &2.94\\
$[$N II] $\lambda$6583                	  &0.71 ($\pm$0.14) &0.62\\
He I$\lambda$6680                         &0.04 ($\pm$0.01) &-\\
$[$S II] $\lambda$6716	                  &0.86 ($\pm$0.09) &0.80\\
$[$S II] $\lambda$6731	                  &0.98 ($\pm$0.10) &0.99\\
He I $\lambda$7065                        &0.07 ($\pm$0.02) &-\\
$[$Ar III] $\lambda$7136		  &0.22 ($\pm$0.03) &-\\
$[$O II] $\lambda$7325			  &0.35 ($\pm$0.05) &0.34\\
$[$Ar III] $\lambda$7751		  &0.06 ($\pm$0.02) &-\\
O I $\lambda$8446			  &0.08 ($\pm$0.02) &-\\
$[$S III] $\lambda$9069 		  &0.48 ($\pm$0.09) &0.42\\
$[$S III] $\lambda$9532                   &2.00 ($\pm$0.28) &1.02\\
\tablebreak
$[$Si IX] 3.9357 $\mu$m~~~~~~~~~~~~~~~    &$<$0.06~~~~~~~~ &-\\
$[$S IV] 10.5105 $\mu$m                    &0.76 ($\pm$0.28) &0.29 \\
$[$O IV] 25.8903 $\mu$m                    &1.90 ($\pm$0.38) &0.41\\
$[$S III] 33.4810 $\mu$m                   &$<$1.24~~~~~~~~  &0.30 \\

\tablenotetext{a}{Flux (narrow H$\beta$) $=$ (2.57 $\pm$0.24) x 10$^{-14}$
ergs s$^{-1}$ cm$^{-2}$.}
\tablenotetext{b}{E$_{B-V}$ $=$ 0.05 $\pm$0.03 mag.}
\tablenotetext{c}{Relative contributions from NLR models: 30\% INNER,
20\% MIDDLE, and 50\% OUTER.}  
\enddata
\end{deluxetable}

\clearpage
\begin{deluxetable}{lrr}
\tablecolumns{3}
\footnotesize
\tablecaption{Comparison of Broad-Line Ratios (relative to broad H$\beta$$^{a}$)
to BLR Model Predictions\label{tbl-5}}
\tablewidth{0pt}
\tablehead{
\colhead{Line} & \colhead{~Observed~} 
& \colhead{~Model$^{b}$}
}
\startdata
Si IV $\lambda$1398 $+$ O IV] $\lambda$1402 &0.80 ($\pm$0.35) &0.98\\
C IV $\lambda$1550          	      	 &4.89 ($\pm$1.06) &4.87\\
He II $\lambda$1640         	      	 &1.12 ($\pm$0.40) &1.07\\
C III] $\lambda$1909                     &1.47 ($\pm$0.37) &1.25\\
Mg II $\lambda$2800                      &0.37 ($\pm$0.14) &0.95\\
H$\delta$ $\lambda$4102                  &0.21 ($\pm$0.07) &0.22\\
H$\gamma$ $\lambda$4340                  &0.43 ($\pm$0.12) &0.47\\
He II $\lambda$4686          	      	 &0.19 ($\pm$0.10) &0.15\\
He I $\lambda$5876           	      	 &0.13 ($\pm$0.04) &0.08\\
H$\alpha$ $\lambda$6563      	      	 &5.13 ($\pm$0.90) &4.09\\
\tablenotetext{a}{Flux (broad H$\beta$) $=$ (2.07 $\pm$0.31) x 10$^{-14}$
ergs s$^{-1}$ cm$^{-2}$.}
\tablenotetext{b}{U $=$ 10$^{-2.25}$, 1/2 solar (1/6 N/H), n$_{H}$ $=$ 
3 x 10$^{10}$ cm$^{-3}$, to N$_{eff}$ $\approx$ 7 x 10$^{20}$ cm$^{-2}$,
flux (H$\beta$ at ionized face) $=$ 2.31 x 10$^{6}$ ergs s$^{-1}$ cm$^{-2}$.}
\enddata
\end{deluxetable}
				       				       
\clearpage
\plotone{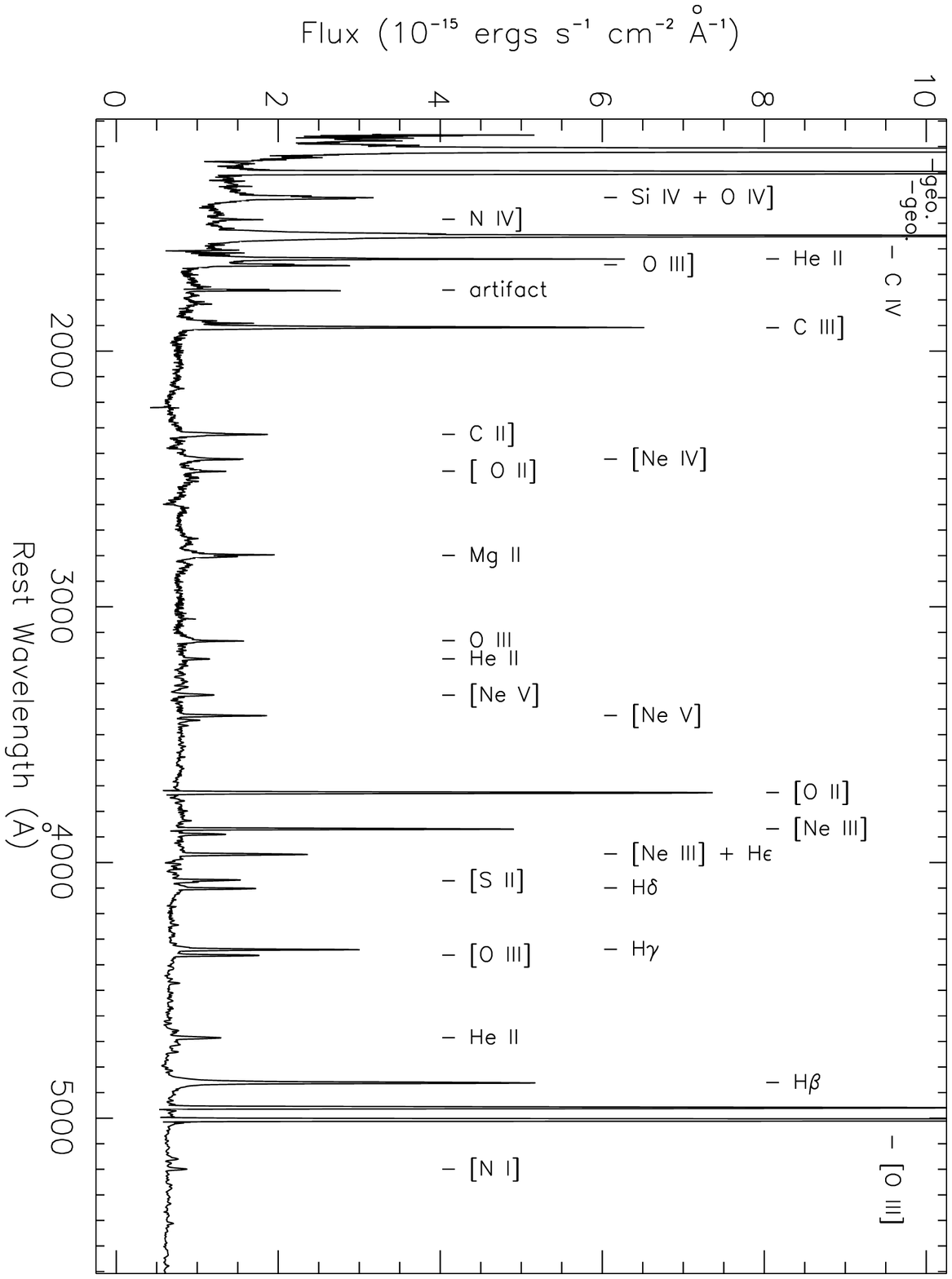}

\clearpage
\plotone{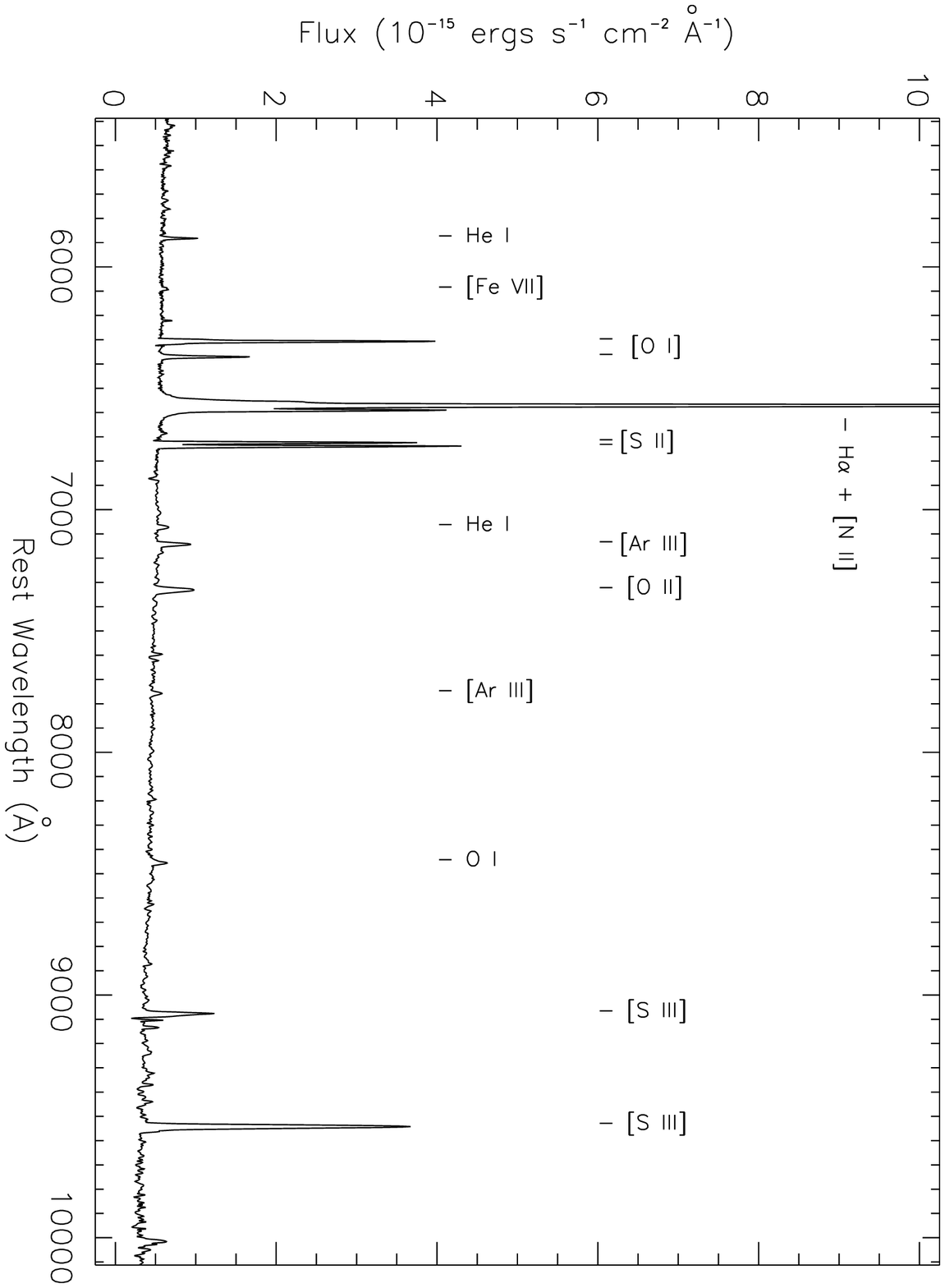}

\clearpage
\plotone{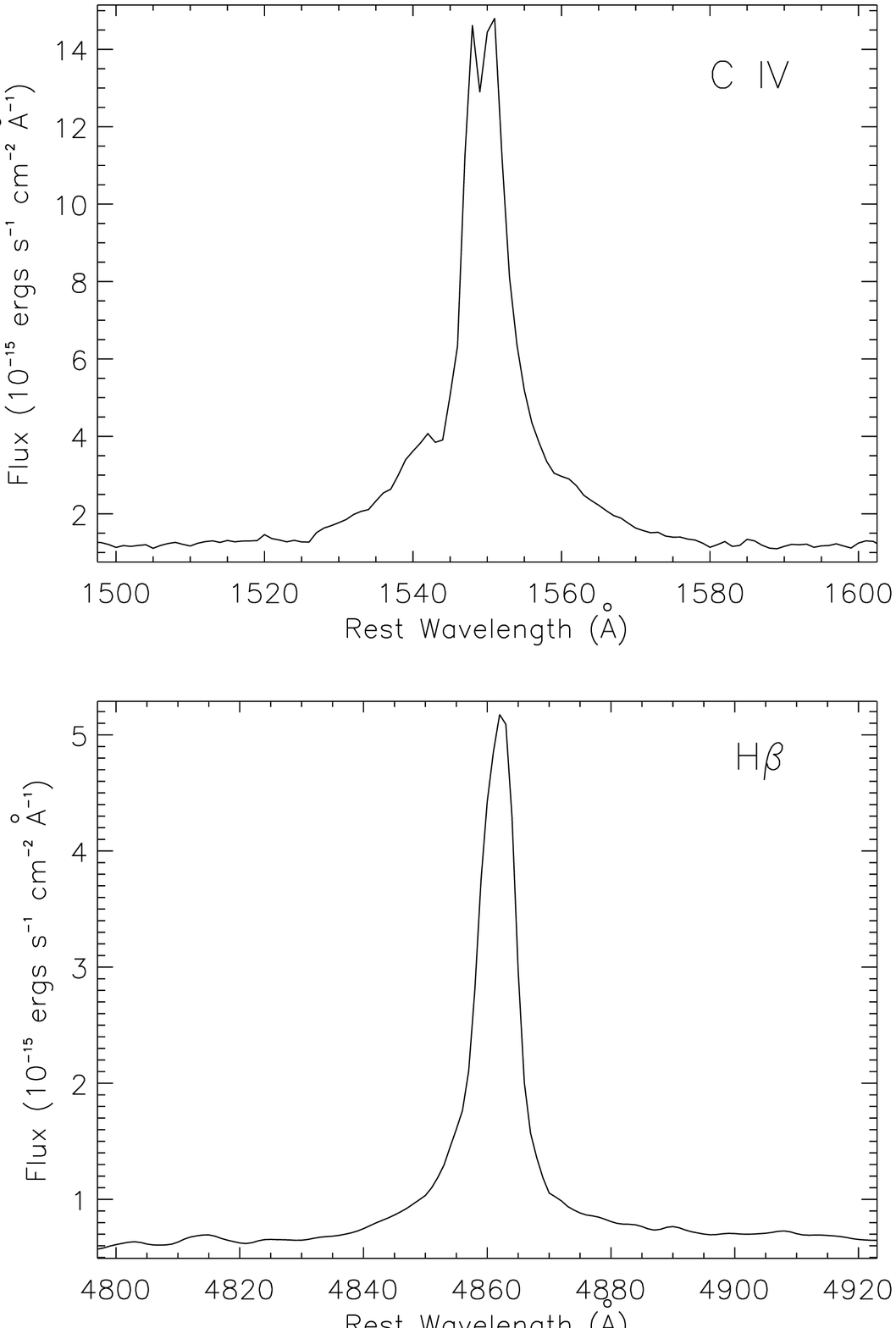}

\clearpage
\plotone{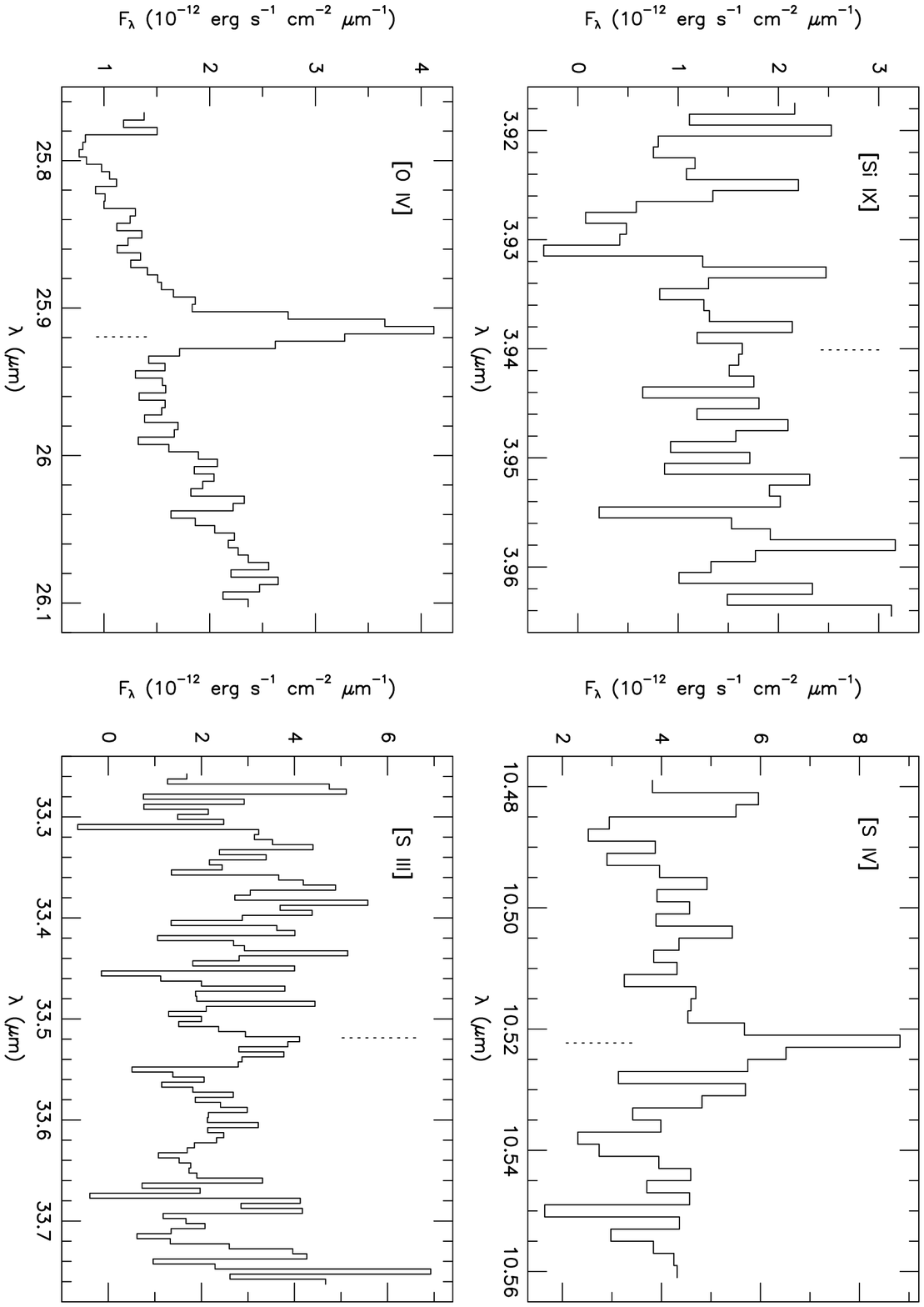}

\clearpage
\plotone{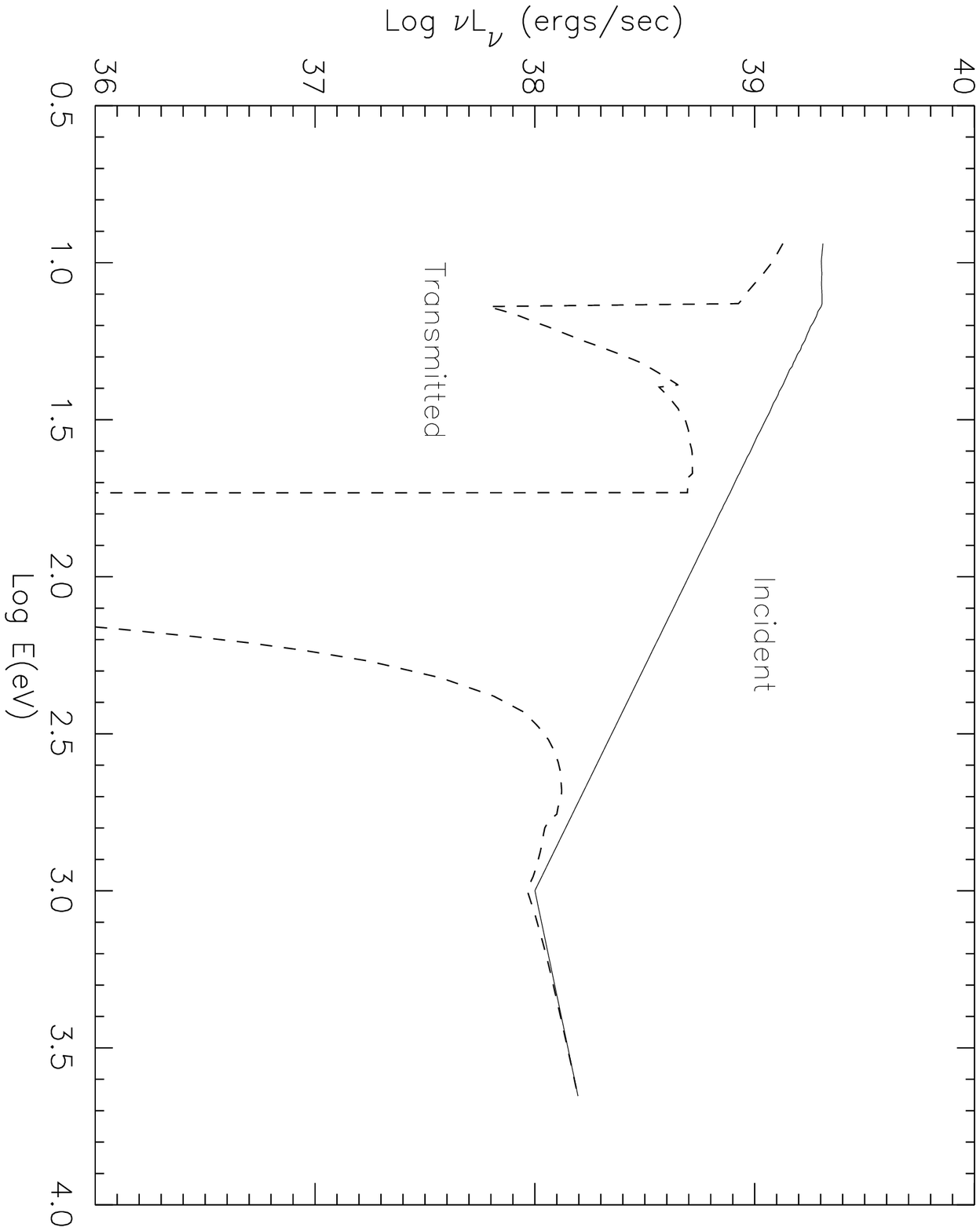}

%\clearpage
%\plotone{fig4.eps}

%\clearpage
%\plotone{fig5.eps}

\end{document}